\documentclass[a4paper]{jpconf}
\usepackage{graphicx}
\begin{document}

\title{Topological view of quantum tunneling coherent destruction}

\author{Alex E. Bernardini and Mariana Chinaglia}
\address{Departamento de F\'{\i}sica, Universidade Federal de S\~ao Carlos, PO Box 676, 13565-905, S\~ao Carlos, SP, Brasil.}
\ead{alexeb@ufscar.br}

\begin{abstract}
Quantum tunneling of the ground and first excited states in a quantum superposition driven by a novel analytical configuration of a double-well (DW) potential is investigated. Symmetric and asymmetric potentials are considered as to support quantum mechanical zero mode and first excited state analytical solutions. Reporting about a symmetry breaking that supports the quantum conversion of a zero-mode stable vacuum into an unstable tachyonic quantum state, two inequivalent topological scenarios are supposed to drive stable tunneling and coherent tunneling destruction respectively. A complete prospect of the Wigner function dynamics, vector field fluxes and the time dependence of stagnation points is obtained for the analytical potentials that support stable and tachyonic modes.
\end{abstract}

\section{Introduction}

\hspace{1 em}The phase-space Wigner-Weyl formalism provides a suitable description of quantum mechanics. In the context of the dynamical evolution of quantum states, the Wigner function helps one to identify highlighting non-classical features revealed, for instance, by quantum tunneling \cite{02,03,01}. The investigation of single-particle tunneling in double-well (DW) potentials can be found in several relevant scenarios \cite{04,05,06,07A,07B,07C,07D}. For DW potentials, the maximum point ($\partial^2 V/\partial s^2 < 0$) between the two minima plays the role of a barrier that drives the quantum tunneling exhibited by the quantum superposition of ground and first excited states, when their energy are smaller than that from the hump.
The quantum tunneling for a function generated by a two-wave superposition supported by novel symmetric and asymmetric DW potential configurations, when described in the phase-space, is investigated in this letter.
In particular the dynamics of tachyonic modes supported by modified topological scenarios \cite{08A,08C}, different from those which holds stable tunneling, is discussed.
A symmetry breaking that supports the quantum conversion of a zero-mode stable vacuum into an unstable tachyonic quantum state is assumed \cite{01} as to identify two inequivalent topological scenarios which drive stable tunneling and coherent tunneling destruction, respectively. In this scenarios, the Wigner flow stagnation points can be found where the potential is force free. They depend deeply on the potential critical points and bring on different effects whether they are a minimum, maximum or saddle points identified by winding numbers defined through \cite{02} $\omega=\frac{1}{2\pi} \int d\varphi$, where $\varphi$ is the angle between the positive position axis and the Wigner flow vectors.
After a complete discussion of the tunneling phenomenon where the Wigner functions and their flow vector fields are computed, one identifies the non-classical features regarding the behavior of the stagnation points. 

To summarize, proceeding is organized as following: in Sec. $2$, the framework for the calculation of the ground and first excited states for DW potential configurations is presented, and results for a kink-like deforming function \cite{08A,08B,08C} are obtained; in Sec. $3$ the quantum tunneling and Wigner flow for quantum superposition involving zero-mode stable and tachyonic states are calculated and discussed for symmetric and asymmetric DW configurations; the conclusions are drawn in Sec. $4$.

\section{Ground and first excited states in a DW potential}

\hspace{1 em} Given the ground state, $\psi_{0}$, of a $DW$ potential, well-stablished quantum mechanical results, as those from \cite{03}, allow one to calculate the corresponding first excited, $\psi_{1}$, state through a multiplier function that relates $\psi_{0}$ with $\psi_{1}$ as \cite{02}
\begin{equation}
\psi_{1}=\alpha\psi_{0}.
\label{mfunction}
\end{equation}
Through the substitution of $\psi_1$ by $\alpha \psi_0$ into the Schr\"{o}dinger equation $\psi''_n(s)+[E_n-V(s)]\psi_n(s)=0$ for $n = 1$ and subtracting the result from the Schr\"{o}dinger equation for $n = 0$ multiplied by $\alpha$, one can find the quantity $\chi$,
\begin{equation}
\chi(s)=\frac{\alpha''+\Delta E \alpha}{2\alpha'}=-\frac{\psi_{0}'}{\psi_{0}}-=-\frac{d \ln\psi_{0}}{ds},
\label{chi}
\end{equation}
where $\Delta E$ is the energy difference between the states and henceforth shall be conveniently denoted by $\Delta E =1/n$.
Considering that $\alpha$ is an {\em ansatz} multiplier function, for a ground state described by,
\begin{equation}
\psi_{0}(s)=\mathcal{N}_0 \exp\bigg( -\int_{0}^{s}\chi(s')ds'  \bigg),
\label{psi0}
\end{equation}
where $\mathcal{N}_0$ is a normalization constant, the first excited state can be calculated through Eq.~(\ref{mfunction}).
Once a solution is found, the potential can be straightforwardly calculated through
\begin{equation}
V(s)=\frac{\psi_{0}^{''}}{\psi_{0}}+E_{0}=\chi^2-\chi'+E_{0}.
\label{potential}
\end{equation}
An {\em ansatz} multiplier function set as $\alpha_1 = \beta+\tanh(s)$ \cite{02,01} which is, up to an additive constant, the solution for the differential equation driven by the $\lambda \phi^4$ model. This multiplier function is an input for both set of potentials: symmetric ($\beta=0$) and asymmetric ($\beta\neq0$) ones. Once one has chosen $\alpha$, it is possible to find $\chi$ through Eq.~(\ref{chi}) and proceed with the calculations for $\psi_0$ and $\psi_1$ by means of Eq.~(\ref{psi0}) and Eq.~(\ref{mfunction}), respectively. The potential can be calculated either by means of $\chi$ or $\psi_0$ through the Eq.~(\ref{potential}).

Once $\psi_0$ and $\psi_1$ have been established, two inequivalent topological scenarios that supports the quantum conversion of a zero-mode stable vacuum into an unstable tachyonic quantum state can be found. As to see the connection between their energies, one assumes a scalar field equation of motion, $\frac{\partial^2\phi}{\partial s^2}=\frac{dU}{d\phi},$ derived from a time independent Lagrangian in $1D$, from which, through quantum perturbations, by an increment of $\xi$, it is possible to find the Schr\"{o}dinger-like equation,
\begin{equation}
\bigg( -\frac{d^2}{ds^2}+V_0^{(QM)}(s) \bigg) \psi_n(s)=w^2_n\psi_n(s),
\label{schlike}
\end{equation}
with $V_0^{(QM)}(s)=d^2U(s)/d\xi^2(s)$, i. e. the quantum mechanical potential analogue. The eigenenergy, when real, i.e. for $w^2>0$, leads to a quantum stable solution. When it is imaginary, i.e. for $w^2<0$, one deals with unstable functions: the tachyonic modes.

By identifying $U(\xi)=\frac{z^2_{\xi}}{2}$, with $z_{\xi}=dz/d\xi=d\xi/ds=\xi^{'}$, it is straightforward to show that Eq.~(\ref{schlike}) can be written as,
\begin{equation}
\bigg( -\frac{d^2}{ds^2}+\xi^{'''}/\xi^{'} \bigg) \psi_n(s)=w^2_n\psi_n(s).
\end{equation}

Note that if the $\psi_1$ energy, $w_1$, is converted to zero, one has a novel scenario where the Schr\"{o}dinger-like equation is
\begin{equation}
\bigg( -\frac{d^2}{ds^2}+V_1^{(QM)}(s) \bigg) \psi_1(s)=0,
\end{equation}
with $V_1^{(QM)}$ the novel quantum potential \cite{01}. Now, there is a state that cross the origin producing a node.
Eq.~(\ref{mfunction}) thus implies that,
\begin{equation}
\bigg( -\frac{d^2}{ds^2}+V_1^{(QM)}(s) \bigg) \psi_0(s)=-w_1^2\psi_0.
\end{equation}

Once one has the zero mode $\psi_0$ calculated through Eq.~(\ref{psi0}) and the potential found in Eq.~(\ref{potential}), it is possible to compare it with $V_0^{(QM)}$ and identify that $\psi_0\propto\xi^{'}$. So there is an identification with the corresponding topological scenario that supports quantum perturbations for this $V_0^{(QM)}$ potential \cite{01}.
In fact, the zero-mode stable vacuum ($\psi_0$) can work as a mathematical tool for obtaining an unstable quantum state, which behaves like a tachyonic mode. Therefore, in this case, $\psi_1$ becomes the novel zero-mode stable solution in the tachyonic system causing the appearance of instabilities, as $\psi_0+e^{-it/n}\psi_1$ is converted into $e^{-t/n}\psi_0+\psi_1$, namely, an unstable configuration.
For a DW potential as in (\ref{potential}) built from the ground state wave function, one identifies
\begin{equation}
\psi_{0} = \mathcal{N}_0 \,2^{1/n} \,\exp{\bigg{(}-\frac{(\beta + 2\arctan(e^s)-\frac{\pi}{2}) \sinh[s]}{2 n}\bigg{)}}
\cosh[s]^{\frac{n+1}{2n}},
\end{equation}
as well as the multiplier function, $\alpha=\beta + 2\arctan(e^s)-\frac{\pi}{2}$,
where $\beta\neq 0$ drives the asymmetric profiles, as obtained from Eq.~(\ref{potential}).
\begin{figure}[b!]
\centering
\includegraphics[scale=.73]{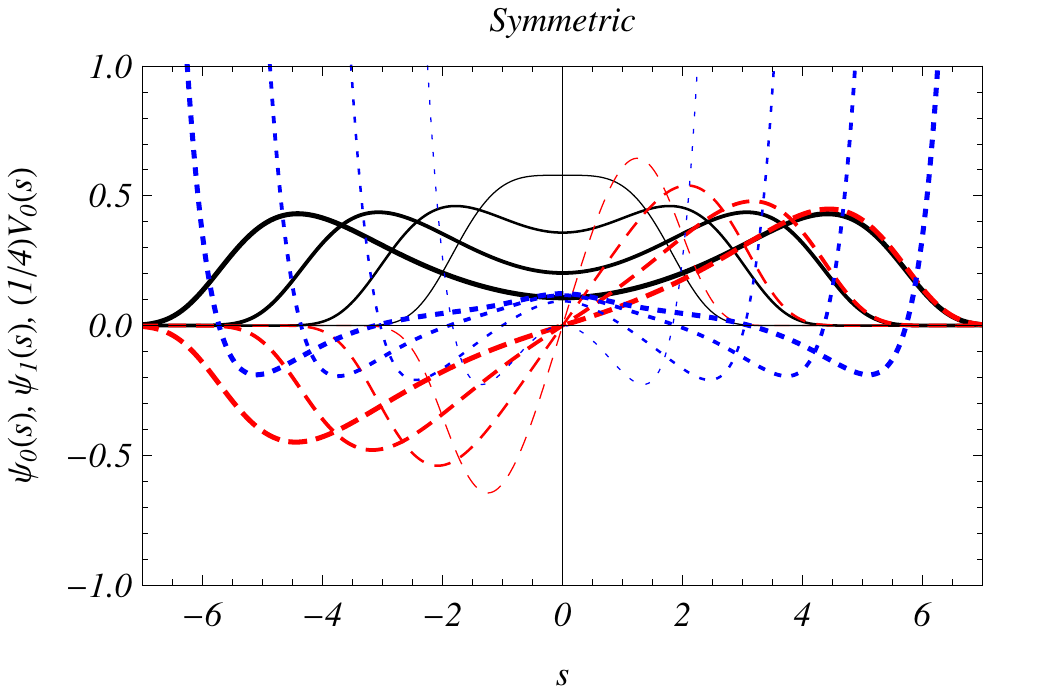}
\includegraphics[scale=.73]{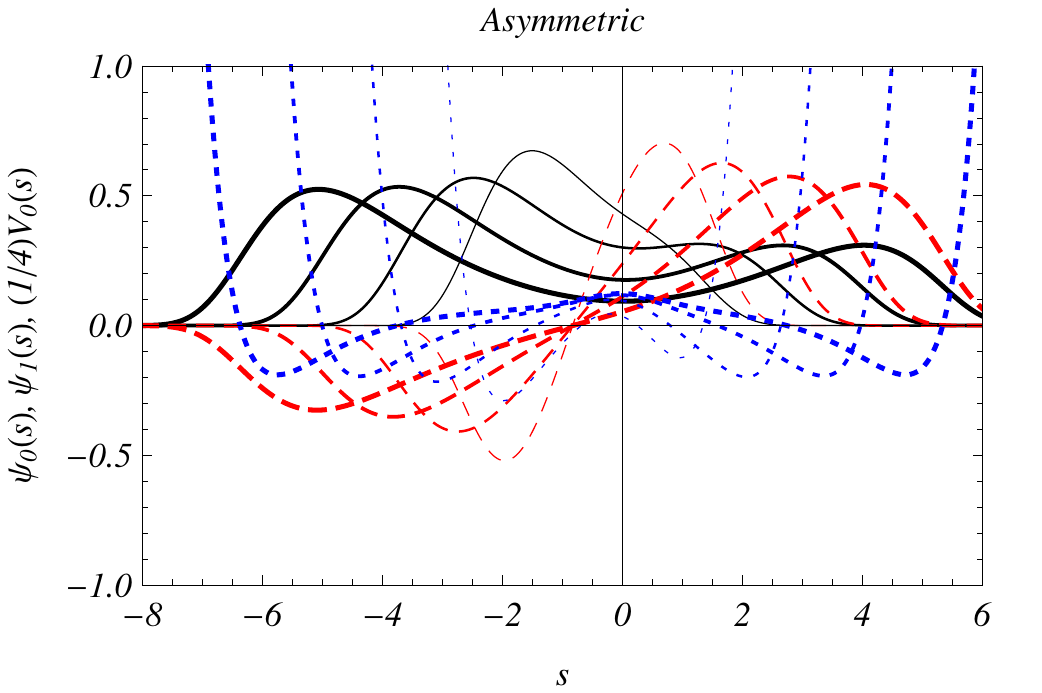}
\caption{(Color online) Ground (solid black lines) and first excited (dashed red lines) states for symmetric ($\beta=0$) (first plot) and asymmetric ($\beta=3/4$) (second plot) QM potentials (dotted blue lines).
The plots are for $\Delta E = 1/n$, with $n = 1$ (thickest), $4,\,16$, and $64$ (thinnest).}
\label{Figcaso3}
\end{figure}
The corresponding first excited states are obtained from Eq.~(\ref{mfunction}), also for symmetric and asymmetric DW configurations. Fig. \ref{Figcaso3} shows the ground and first excited states along with both potentials, symmetric and asymmetric ones.

\section{Quantum tunneling and Wigner flow analysis}

\begin{figure}[b!]
\centering
\includegraphics[scale=.73]{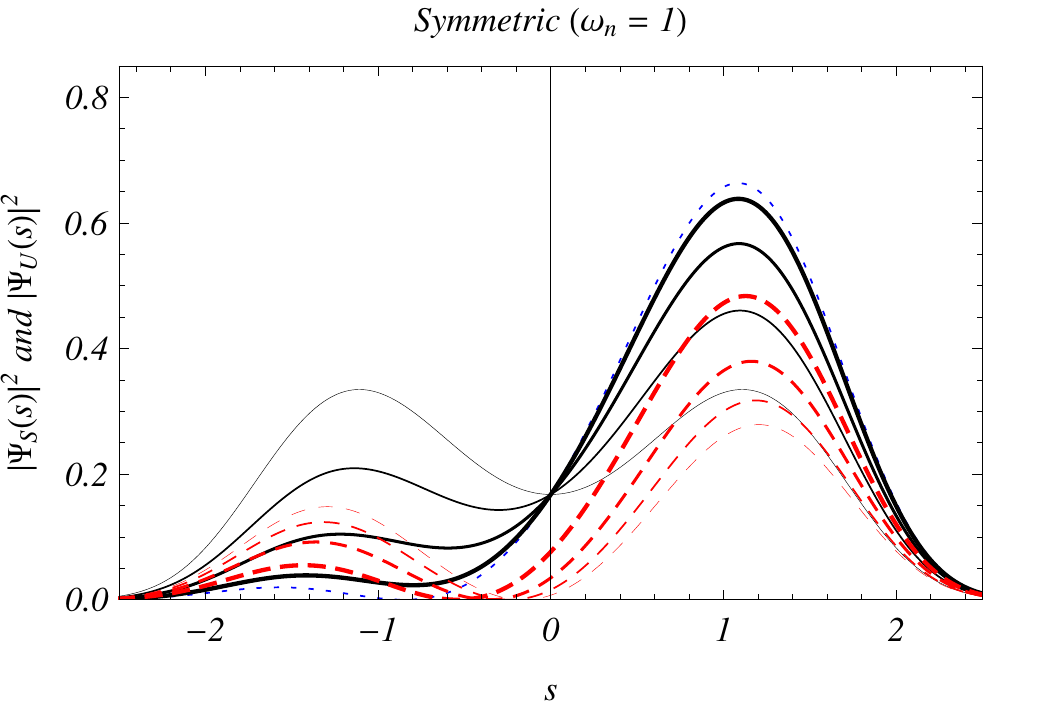}
\includegraphics[scale=.73]{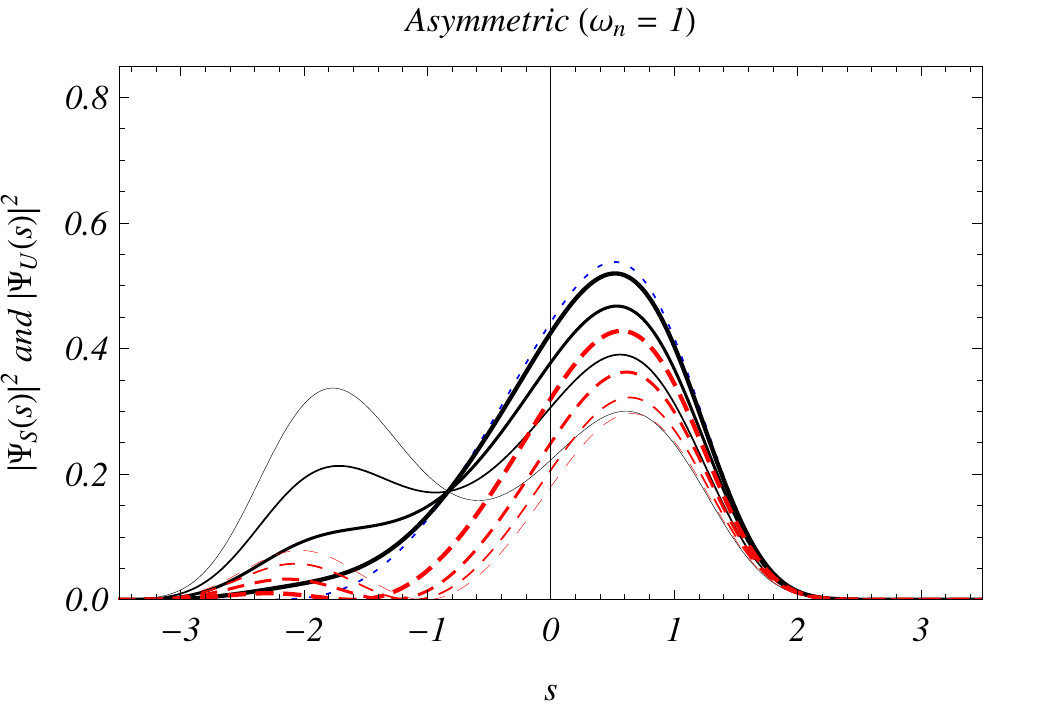}
\includegraphics[scale=.73]{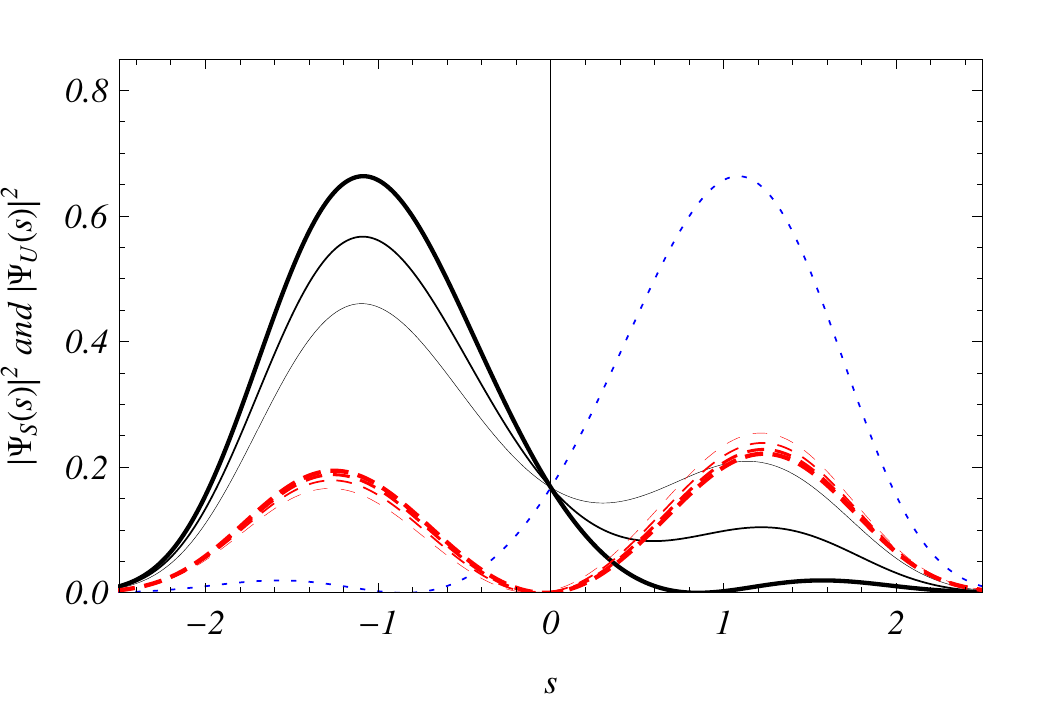}
\includegraphics[scale=.73]{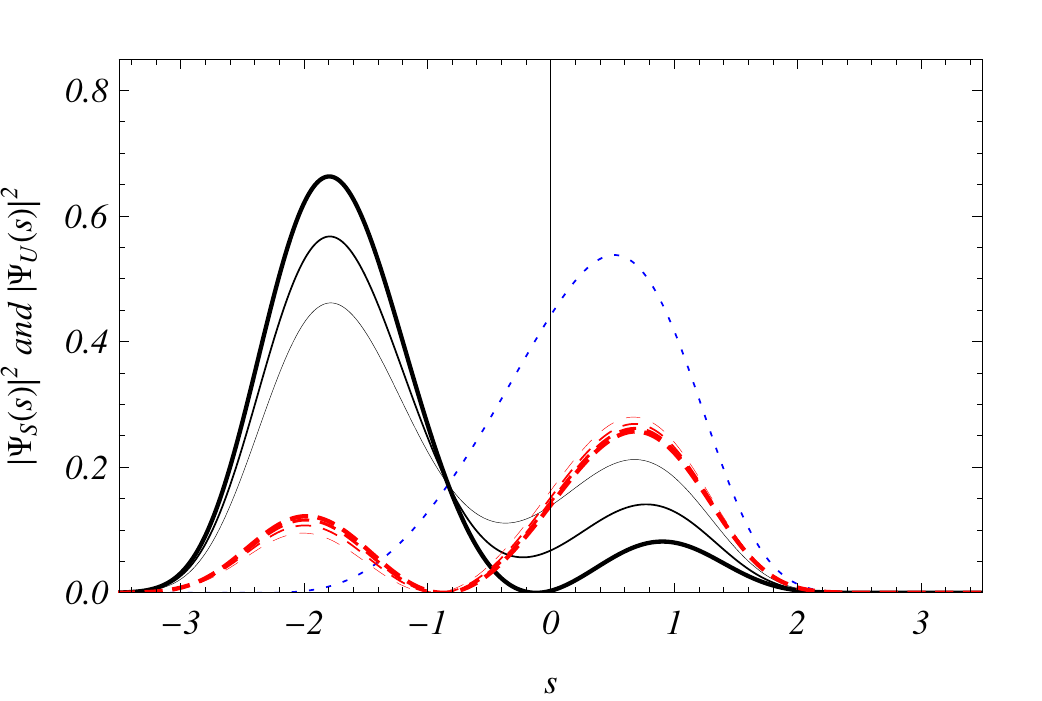}
\caption{
(Color online) Time evolution of probability densities for stable (solid black lines) and unstable (dashed red lines) composite states, $\Psi_S$ and $\Psi_U$, in DW symmetric (first row) and asymmetric (second row) configuration.
The curves are for times running from $t = \pi/8$ (thickest line) to $t = \pi/2$ (thinnest line) in the first plot, and from $t = 5\pi/8$ (thinnest line) to $t = \pi$ (thickest line) in the second plot, with steps $\Delta t = \pi/8$.
Dotted blue lines correspond to $t = 0$ for both, stable and unstable tunneling.
The results are for $n = 1$.}
\label{Compare3}
\end{figure}

\begin{figure}
\centering
($a$)\includegraphics[scale=.58]{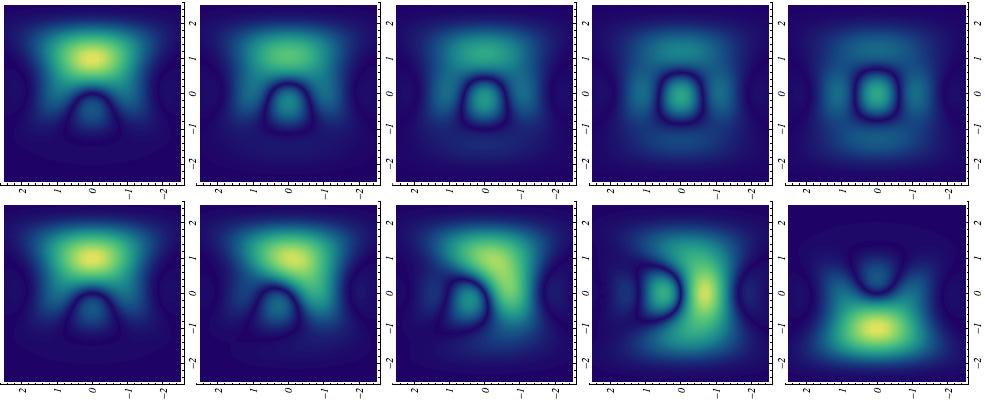}
($b$)\includegraphics[scale=.58]{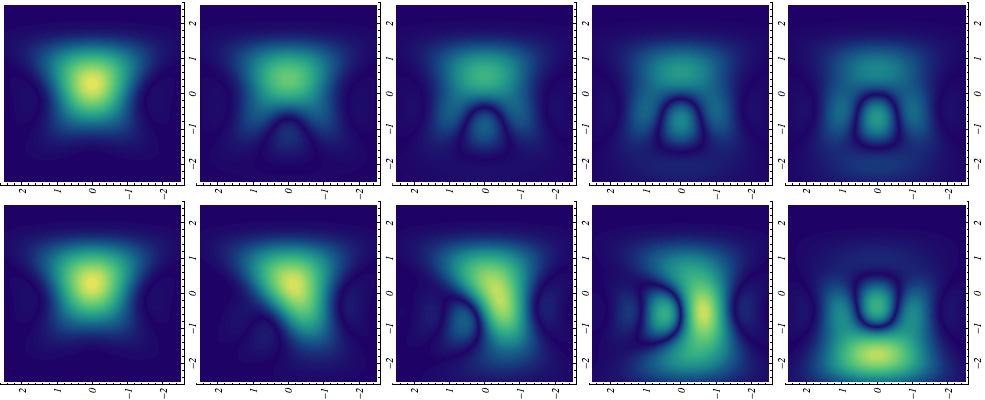}
\caption{(Color online)Time evolution of Wigner amplitudes relative to symmetric ($a$) and asymmetric ($b$), unstable (first and third rows) and stable (second and forth rows) wave function configurations.
The pictures are shown for time evolution from left ($t=0$) to right ($t = \frac{\pi}{2}$) for time intervals of $\frac{\pi}{8}$. Here again it was considered $n=1$. Vertical and horizontal axes measure positions and momenta respectively.}
\label{Wigner3}
\end{figure}

\hspace{1 em}In order to investigate the quantum tunneling one needs the superposition function of the ground, $\psi_0$, and first excited, $\psi_1$, states:
\begin{equation}
\Psi(s,t)=\sin(\theta)exp \bigg(\frac{-iE_0t}{\hbar}\bigg)\psi_0(s)+\cos(\theta)exp \bigg(\frac{-iE_1t}{\hbar}\bigg)\psi_1(s),
\label{psigrande}
\end{equation}
which will be set into the Wigner function:
\begin{equation}
W(s,p,t)=\frac{1}{\pi \hbar}\int_{- \infty}^{\infty} dy\Psi^*(s+y,t)\Psi(s-y,t)e^{\frac{2i}{\hbar}py},
\end{equation}
where $\theta$ denotes the weight of $\psi_0$ and $\psi_1$ in the superposition function. As a quasi-probability function, the Wigner function attains only real values they can also be negative. The modulus of the normalized stable-unitary (black lines), and unstable$\rightarrow$stationary (red lines) density probabilities are depicted in Fig.~\ref{Compare3} for different times, ranging from $t=0$ to $t=\pi/2$ (first row) and from $t=5\pi/8$ to $t=\pi$ (second row).
It is evidenced through the images that the normalized function completes the tunneling cycle going from {\em right to left} for both potentials while the tachyonic cycle is converted into a stationary state after the coherent destruction of the unstable mode.
The Wigner functions squared for both scenarios relative to symmetric and asymmetric potentials can be seen in Fig. \ref{Wigner3}. Analyzing the image it is clear, as was previously seen with Fig.~\ref{Compare3}, that the tunneling occurs for both functions. The difference appears when one recognizes that for the normalized wave function the tunneling process is invertible whereas for tachyonic functions a limit state is reached.
The invertible behavior appears because of the small difference between the energy of the states that causes a beat period, $T=2\pi\hbar/\Delta E$. The complete beating, that is when the function goes from one well to another and returns, is not presented in the tachyonic case because it is not normalizable. This latter smoothly attains an equilibrium lying in equal proportions in the two wells.
The next step corresponds to the calculation of the Wigner flow,
\begin{equation}
\left( \begin{array}{c}
J_s \\ \\
J_p \end{array} \right)=\left( \begin{array}{c}
\frac{p}{m}W(s,p,t) \\ \\
-\sum_{l=0}^{l=\infty} \frac{(\frac{i\hbar}{2})^{2l}}{(2l+1)!}\frac{\partial^{2l}W(s,p,t)}{\partial_{p}^{2l}}\frac{\partial^{2l+1}V(s)}{\partial^{2l+1}s} \end{array} \right).
\end{equation}
which, in the classical limit, i.e. when one has $\hbar\rightarrow0$ or $V(s)$ powers $\leq2$, leads to an equivalence between Wigner and classical Liouville flow.
\begin{figure}
\vspace{-5cm}[b!]
\centering
\includegraphics[scale=.78]{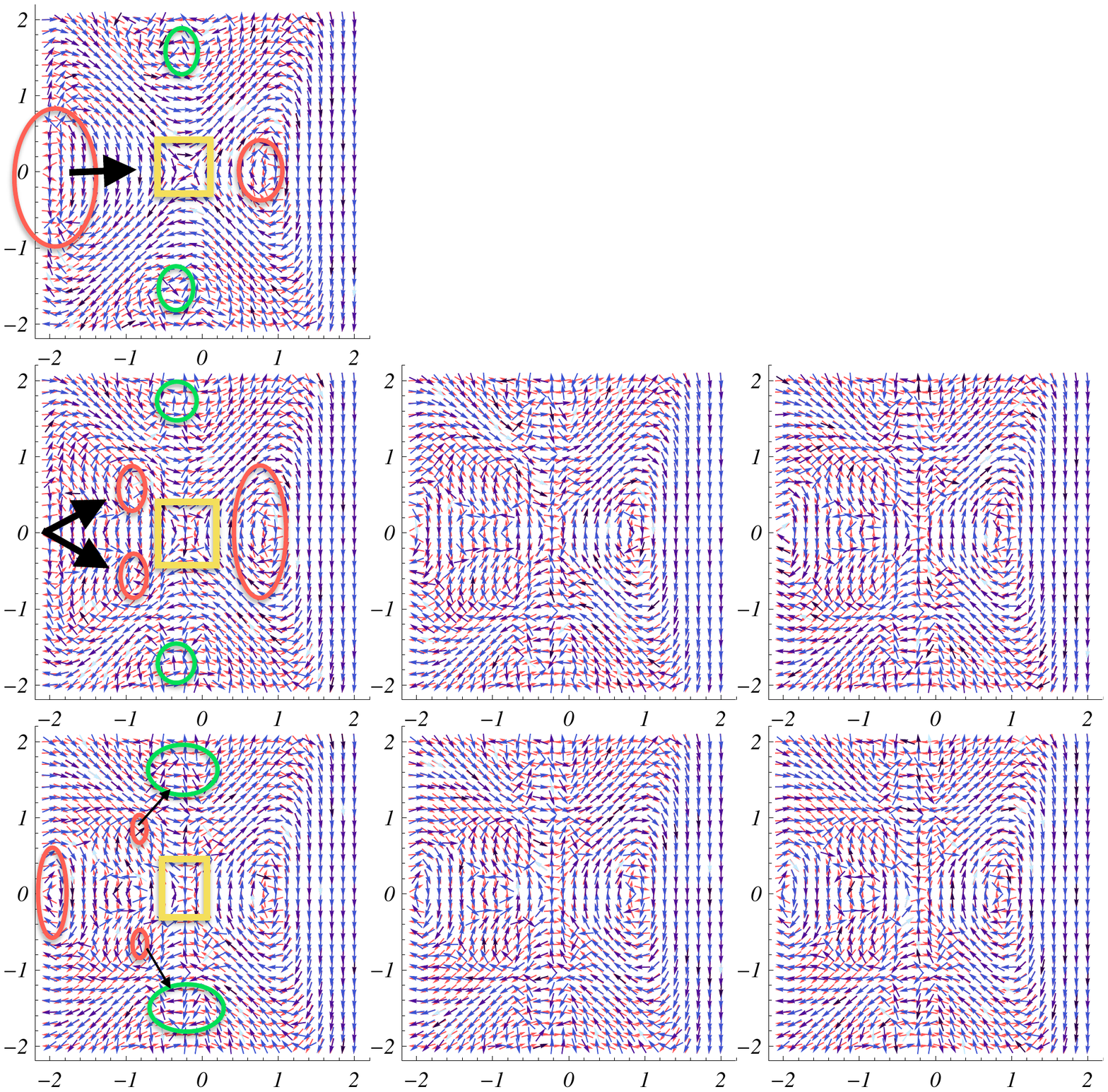}
\vspace{-4cm}
\caption{(Color online) First line portrait corresponds to $t=0$. Last two lines represent the time evolution of the stable wave function for the asymmetric potential (last row of Fig.~3) up to half tunneling period. Red and green circles represent a CW and CCW vortex respectively, therefore both contributes with $\omega=+1$. Yellow rectangle represents a saddle point and contributes with $\omega=-1$. Vertical and horizontal axes measure $p$ and $x$, respectively}
\label{Vec31}
\end{figure}
By convenience, the Wigner flow vector fields relative to the stable wave functions driven by the asymmetric potential is considered in Fig.~\ref{Vec31} for different times along the tunneling evolution. The first feature that is highlighted through the vector field is the difference between the classical (red) and quantum (blue) vector directions. The larger this difference, the greater the degree of non-classicality of the system \cite{02}.
While passing through a minimum of the potential, the flow spins towards this point and one has the winding number $\omega=+1$. Passing through a maximum of the potential, the flow also spins generating $\omega=+1$, but it spreads instead of concentrating. If it is a saddle point it elongates the flow making it slower and slower near the central point, through which one has $\omega=-1$.

Another feature is the flux reversion that corresponds to the regions where the Wigner function attains negative values. In order to investigate this feature ally with the stagnation points position, one can imagine a top-down vertical line around $s=-0.25$ in the first frame of Fig.~\ref{Vec31}. Following this line it is possible to first recognize one counterclockwise (CCW) vortex (green circle), then a flow reversal followed by a saddle point (yellow square), around $p=0$. Moving to negative values of $p$ one has a CCW vortex, causing again the flow inversion. This behavior is a further indication of the non-classical character of the system.

In order to see the stagnation points dynamics and calculate $\omega$, the integration loop is depicted in each frame of Fig. \ref{Vec31}. Note that the initial total winding number (first frame) is $\omega=+2$, correspondent to one clockwise (CW) vortex (red circle; $\omega=+1$), two CCW vortices (green circles; $\omega=+1$) and one saddle point (yellow square); $\omega=-1$). In the next frame the saddle point starts to crumble while another CW vortex appears in the left. Besides, one can no longer recognize the CCW vortices, which results again in a total winding number $\omega=+2$. Considering all subsequent frames one sees that the saddle point completely disappears engendering sequentially $\omega=+2$. Consequently, charge conservation is attained as conjectured \cite{02}. The novelty is that an opposite conclusion appears when considering the unstable case for which one has no charge conservation.
For the unstable function however this lack of conservation is not totally a surprise, once it is predicted only for systems with unitary quantum dynamics.
Finally, the stagnation points also work as a tool for the recognition of tunneling. The function $\Psi$, which is found initially only in the right minimum of the potential, appears in the form of a CW vortex near $s=0.8$ in Fig.~\ref{Vec31}. After time evolution $\Psi$ coexists in both minima, thus causing the appearance of another CW vortex in the left-side in some subsequent frames. This behavior reveals the tunneling through the stagnation points.

\section{Conclusions}

\hspace{1 em} Quantum tunneling analysis has shown that one has a reversibility of the process for the stable wave function whereas the tachyonic function attains a limit coexisting in both wells for any subsequent time. Thereat one can say that the tunneling undergoes a process of coherent destruction.
It was also noticed that the Wigner vector field can be used to study non-classical features since it evinces the differences between quantum and classical vector flux profiles as well as the flow reversibility. The latter occurs where one has negative values for the Wigner function related to the stagnation points. Lastly, while one finds the expected charge conservation for stable wave functions, one draws attention to the fact that this conservation is not in principle expected for unstable functions. In fact, it is depicted from the vector flow analysis for the tachyonic case.

\ack{AEB thanks for the financial support from the Brazilian Agency FAPESP (grant 15/05903-4).}

\smallskip
\end{document}